\documentclass[%
 aip, apl,
 sd,%
 amsmath,amssymb,
reprint,%
]{revtex4-1}

\usepackage{graphicx}
\usepackage{dcolumn}
\usepackage{bm}

\newcommand{\bhline}[1]{\noalign{\hrule height #1}}

\draft 
\begin{document}


\title{Particle-scale modeling of the drying characteristics of colloidal suspensions} 



\author{Rei Tatsumi}
  \email{tatsumi@sogo.t.u-tokyo.ac.jp}
\affiliation{ 
Environmental Science Center, The University of Tokyo, Tokyo 113-8656, Japan
}
\author{Osamu Koike}
\author{Yukio Yamaguchi}
\affiliation{
Product Innovation Association, Tokyo 113-8656, Japan
}
\author{Yoshiko Tsuji$^{1,}$}
\affiliation{ 
Department of Chemical System Engineering, The University of Tokyo, Bunkyo, Tokyo 113-8656, Japan
}

\date{\today}

\begin{abstract}
During drying of colloidal suspensions, 
colloidal particles can form concentrated particle layers beneath the receding free surface.
The drying rate can gradually decrease with the growth of the particle layers. 
We construct a model to investigate how such drying characteristics is affected by interactions between particles.
In this model, the formation of the particle layers is described by Langevin dynamics simulations,
and the drying rate is evaluated from the permeation resistance of the particle layers.
We show that the decrease in the drying rate is suppressed when the particles form aggregates by attractive interactions.
The present model would enable us to predict and control the drying characteristics through the character of colloidal particles.
\end{abstract}


\maketitle 


\section{\label{s1}Introduction}

Colloidal suspensions are coated on substrates and upon drying produce various functional materials,
whose quality is determined by the microstructure composed of the colloidal particles. 
During drying of colloidal suspensions,
the receding free surface (the liquid--air interface) induces the particles to form structures, 
thereby causing a decrease in the drying rate.~\cite{dryc1, dryc2, dryc3}
The prediction and control of such drying characteristics is therefore required to improve material quality as well as to reduce drying time.

\begin{figure}
 \centering
 \includegraphics[width=8cm]{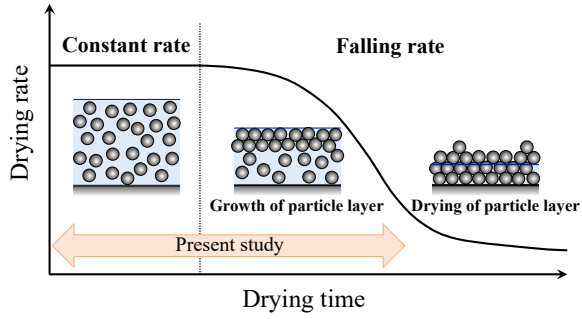}
 \caption{
Sketch of the drying characteristics of colloidal suspensions.
}
 \label{f1}
\end{figure}

Figure~\ref{f1} schematically illustrates the structure formation corresponding to the drying characteristics of colloidal suspensions.
During the constant rate period, the drying rate is controlled by vapor diffusion from the free surface where the liquid evaporates.
The receding free surface sweeps the particles to form concentrated particle layers,
and then the falling rate period begins.
The condition that this structure formation occurs can be quantified by the particle drying P\'{e}clet number as $\mathrm{Pe} = \tau_\mathrm{D} / \tau_\mathrm{E} > 1$, 
where $\tau_\mathrm{D}$ and $\tau_\mathrm{E}$ are the time scales of particle diffusion by the Brownian motion and the recession of the free surface, respectively.~\cite{clyr1, clyr2, sdry1}
The continuous evaporation requires liquid transport through the pores of the particle layer to the free surface.
With the growth of the particle layer, the permeation resistance of the particle layer increases.
The drying rate is thus controlled by the liquid transport through the particle layer and decreases gradually.
Such decrease in the drying rate has been quantitatively observed in experiments of the unidirectional drying of colloidal suspensions confined in a thin rectangular cell.
In these experiments, the drying rate can be evaluated by observing the decrease in the volume of the suspension in the drying cell.~\cite{udry1, udry2, udry3, udry4, udry5, udry6}
When the particle layer recedes with the free surface and grows to reach the substrate with further drying, 
the free surface goes into the particle layer. 
The drying rate is then controlled by vapor diffusion through the particle layer and will further decrease,
as observed for granular materials.
In this stage, liquid flow can be driven by a capillary pressure due to menisci formed among particles.
This liquid transport would suppress the decrease in the drying rate.~\cite{pors1, pors2}

We focus on the decrease in the drying rate due to the growth of concentrated particle layers.
The character of colloidal particles would affect the structure of the particle layers.
For example, in the drying of droplets of colloidal suspensions, 
the morphology of the dried grains changes depending on interactions between the particles.~\cite{sdry2, sdry3}
As for the drying of binary colloidal mixtures containing particles of different sizes,
segregation of smaller particles to the top surface can occur depending on the mixing ratio, the particle size ratio, and drying rates.~\cite{seg1, seg2, seg3}
Such structural differences would be reflected in the drying characteristics.

In this study, we construct a model to investigate how the drying characteristics of colloidal suspensions is affected by interactions between the colloidal particles.
We first derive an analytical expression of the relationship between the drying rate and the permeation resistance of concentrated particle layers.
We then perform Langevin dynamics simulations that describe the Brownian motion of colloidal particles by stochastic differential equations.
The receding free surface is also considered in the simulations to induce the formation of concentrated particle layers.
During the simulations, the drying rate is evaluated from the permeation resistance of the particle layers by use of the derived analytical expression.
We assume the DLVO potential between charged colloidal particles and tune the potential from repulsive to attractive by varying ionic strength.
We show that the decrease in the drying rate is suppressed when the potential energy barrier is so low that the particles form aggregates.
We discuss how the interaction between particles affect the drying characteristics in terms of the structural differences of the particle layers.

\section{\label{s2}Model}

\subsection{\label{s2-1}Drying rates}

\begin{figure}
 \centering
 \includegraphics[height=6cm]{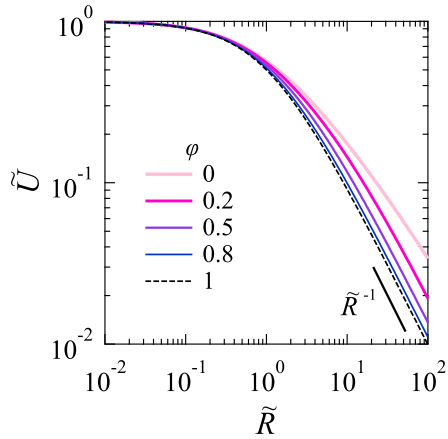}
 \caption{
Drying rate as a function of permeation resistance at different humidities.
The profiles are given by Eq.~(\ref{e2-1-4}) for $\varphi < 1$
and Eq.~(\ref{e2-1-11}) for $\varphi \rightarrow 1$.
}
 \label{f2}
\end{figure}

\begin{figure}
 \centering
 \includegraphics[height=5.5cm]{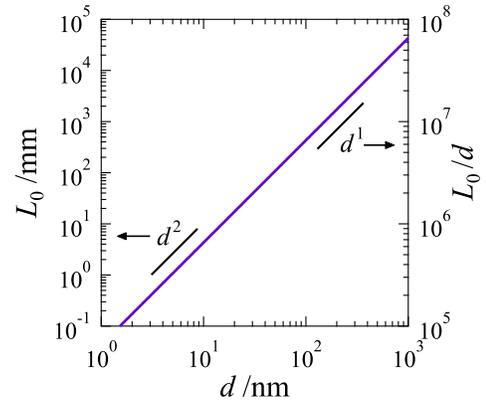}
 \caption{
Characteristic thickness of concentrated particle layers as a function of particle diameter.
The values scaled by $1 \ \mathrm{mm}$ and the particle diameter $d$ are exhibited in the left and right axes, respectively.
The characteristic thickness $L_0 = R_0 / \rho_\mathrm{c}$ is calculated by Eqs.~(\ref{e2-1-7}) and (\ref{e2-1-12}) with setting the parameters:
$\eta = 1 \times 10^{-3} \ \mathrm{Pa \ s}$, $T = 300 \ \mathrm{K}$, $V_\mathrm{m} = 3 \times 10^{-29} \ \mathrm{m^3}$,
$p_0 = 3.5 \times 10^{3} \ \mathrm{Pa}$, $K = 8 \times 10^{-2} \ \mathrm{m/s}$, and $\Phi_\mathrm{c} = 0.64$.
From Eq.~(\ref{e2-1-6}), this parameter setting corresponds to $U_0 = 1 \times 10^{-6} \ \mathrm{m/s}$ for $\varphi = 0.5$.
}
 \label{f3}
\end{figure}

The drying rate $U$ represents the decrease rate of liquid due to evaporation
and equals the mass transfer rate of the vapor from the free surface as
\begin{align}
U = \frac{KV_\mathrm{m}}{k_\mathrm{B} T}(p - p_\infty)
\label{e2-1-1},
\end{align}
where $K$ is the mass transfer coefficient of the vapor, $V_\mathrm{m}$ is the molecular volume of the liquid,
$T$ is the thermodynamic temperature, $k_\mathrm{B}$ is the Boltzmann constant.
Equation~(\ref{e2-1-1}) describes the diffusion of the vapor from the free surface to the ambient air stream.
The stagnant region in vicinity to the free surface is considered as the boundary layer
where the vapor diffuses.
Based on Fick's law,
the mass transfer rate is proportional to the difference in the vapor pressure of saturation $p$ and that of the ambient air $p_\infty$.

We then consider the drying of colloidal suspensions
where concentrated particle layers form as illustrated in Fig.~\ref{f1}.
Evaporation continues by liquid transport through the particle layer to the free surface.
This transport flow is originated by capillarity due to the curvature of the menisci formed among particles.
The permeation resistance of the particle layer induces a pressure drop $\Delta P$,
which is balanced by the capillary pressure. 
Due to the pressure drop, 
the vapor pressure at the free surface decreases from $p_0$ as
\begin{align}
p = p_0 \exp \left( -\frac{V_\mathrm{m} \Delta P}{k_\mathrm{B}T} \right)
\label{e2-1-2},
\end{align}
where $V_\mathrm{m} \Delta P$ gives the chemical potential drop of the liquid.
Equation~(\ref{e2-1-2}) is known as the Kelvin equation.
Since the liquid flow rate equals the drying rate due to mass balance,
the pressure drop is related to the drying rate by Darcy's law as
\begin{align}
\Delta P  = \eta R U
\label{e2-1-3},
\end{align}
where $R$ is the permeation resistance of the particle layer, and $\eta$ is the liquid viscosity.
Combining Eqs.~(\ref{e2-1-1}) -- (\ref{e2-1-3}) yields an equation that relates the drying rate to the permeation resistance as
\begin{align}
\tilde{U} = \frac{e^{-(1 - \varphi) \tilde{R} \tilde{U}} - \varphi}{1 - \varphi}
\label{e2-1-4},
\end{align}
where the relative humidity is given by
\begin{align}
\varphi = \frac{p_\infty}{p_0}
\label{e2-1-5}.
\end{align}
In Eq.~(\ref{e2-1-4}), the drying rate and the permeation resistance are given as dimensionless quantities.
The drying rate $U$ is scaled by that without particle layers $U_0$ as
\begin{align}
\tilde{U} = \frac{U}{U_0}, \ \ \ \ \
U_0 = \frac{K V_\mathrm{m}}{k_\mathrm{B} T}(p_0 - p_\infty)
\label{e2-1-6}.
\end{align}
The permeation resistance $R$ is scaled by a characteristic resistance $R_0$ defined as
\begin{align}
\tilde{R} = \frac{R}{R_0}, \ \ \ \ \
R_0 = \frac{(k_\mathrm{B} T)^2}{\eta K V_\mathrm{m}^2 p_0}
= \frac{1}{\eta U_0} \frac{k_\mathrm{B} T}{V_\mathrm{m}}(1 - \varphi)
\label{e2-1-7}.
\end{align}
As compared to Eq.~(\ref{e2-1-3}),
the definition of $R_0$ indicates that
the flow rate through the particle layer with a permeation resistance of $R_0$ equals $U_0$ 
when a pressure of $(k_\mathrm{B} T/V_\mathrm{m}) (1 - \varphi)$ is applied.
In the previous study, Eq.~(\ref{e2-1-4}) was derived with ignoring humidity as $\varphi = 0$
and was approximately solved with assuming $V_\mathrm{m} \Delta P/ (k_\mathrm{B}T) \ll 1$,~\cite{udry5}
although this assumption only holds for $\tilde{R} \ll 1$.

The solution of Eq.~(\ref{e2-1-4}) is expressed by the Lambert $W$ function that satisfies $x = W(x e^x)$ as
\begin{align}
\tilde{U} = \frac{1}{1 - \varphi} \left[ \frac{1}{\tilde{R}} W(\tilde{R} e^{\varphi \tilde{R}}) - \varphi \right]
\label{e2-1-8}.
\end{align}
As described in Fig.~\ref{f2},
the drying rate decreases with an increase in the permeation resistance.
The asymptotic behavior is different depending on the humidity.
The asymptotic form for $\varphi > 0$ decreases proportional to $\tilde{R}^{-1}$ as
\begin{align}
\tilde{U} \sim -\frac{\ln \varphi}{1 - \varphi} \frac{1}{\tilde{R}}
\ \ \ \ (\tilde{R} \rightarrow \infty)
\label{e2-1-9},
\end{align}
while that for $\varphi = 0$ decreases more slowly as 
\begin{align}
\tilde{U} \sim \frac{\ln \tilde{R}}{\tilde{R}}
\ \ \ \ (\tilde{R} \rightarrow \infty)
\label{e2-1-10}.
\end{align}
Combining Eqs.~(\ref{e2-1-3}) and (\ref{e2-1-9}) shows that
the pressure drop $\Delta P$ approaches to $-(k_\mathrm{B} T/V_\mathrm{m})  \ln \varphi$ as $\tilde{R} \rightarrow \infty$.
If this limiting pressure drop is larger than the maximum capillary pressure achieved by the maximum meniscus curvature,
the free surface will invade into the particle layer during drying.~\cite{cap}

Equation~(\ref{e2-1-8}) becomes a simple expression in the limit of $\varphi \rightarrow 1$ as
\begin{align}
\lim_{\varphi \rightarrow 1} \tilde{U}  = \frac{1}{1 + \tilde{R}}
\label{e2-1-11}.
\end{align}
This expression indicates that the drying rate equals the rate of the liquid flow through the series resistance of $R$ and $R_0$ 
when a pressure of $(k_\mathrm{B} T/V_\mathrm{m}) (1 - \varphi)$ is applied. 
From this expression, the characteristic resistance $R_0$ can be interpreted as the resistance of vapor transport in the boundary layer.
As noted in \ref{sA-2},
Eq.~(\ref{e2-1-11}) has the same form as the Ruth's filtration equation that describes the temporal variation of the flow rate in dead-end filtration.
The Ruth's filtration equation consider the series permeation resistance of the filter media and the cake deposited onto the filter.~\cite{filt}

When we assume that the particle layer is spatially uniform,
the permeation resistance is proportional to the thickness of the layer $L$ as $R = \rho_\mathrm{c} L$
with a constant resistivity $\rho_\mathrm{c}$.
We define a characteristic thickness as $L_0 = R_0/\rho_\mathrm{c}$.
The drying rate will decrease significantly when the particle layer grows as $L > L_0$.
The permeation resistivity can be evaluated by the Kozeny--Carman equation given by
\begin{align}
\rho_\mathrm{c} = \frac{180}{d^2} \frac{\Phi^2_\mathrm{c}}{(1 - \Phi_\mathrm{c})^3}
\label{e2-1-12},
\end{align}
where $d$ is the particle diameter, and $\Phi_\mathrm{c}$ is the volume fraction of the particles in the particle layer.
With assuming the uniformity of the particle layer,
the present model gives an analytical expression of the drying characteristics as derived in \ref{sA-2}.
As shown in Fig.~\ref{f3}, the characteristic thickness $L_0$ increases proportional to $d^2$.
We thus expect that the decrease in the drying rate will not be obvious for large particles.
In fact, in the previous experiments of unidirectional drying in confined systems,
the decrease in the drying rate was observed only for the particles of $d \lesssim 40 \ \mathrm{nm}$.~\cite{udry1, udry2, udry3, udry4, udry5, udry6}
The condition that the drying rate will obviously decrease during drying can be estimated as
$(\Phi_0 / \Phi_\mathrm{c}) H_0 > L_0$,
where $\Phi_0$ and $H_0$ are the initial particle volume fraction and thickness of the coating film of the colloidal suspension, respectively.
In this estimation, $(\Phi_0 / \Phi_\mathrm{c}) H_0$ equals the particle layer thickness when all the particles in the suspension form the particle layer.
When the particles of $d = 10 \ \mathrm{nm}$ form a particle layer with $\Phi_\mathrm{c} = 0.64$ (random close packing~\cite{rcp}), 
the characteristic thickness is $L_0 = 4.3 \ \mathrm{mm}$.
The decrease in the drying rate will be obvious when the initial film thickness satisfies $H_0 > 27.5 \ \mathrm{mm}$ for $\Phi_0 = 0.1$.

\subsection{\label{s2-2}Formation of concentrated particle layers}

We consider a suspension containing spherical colloidal particles of a diameter $d$ and a mass $m$.
The suspension is coated on a plane substrate where we set the $z$-coordinate along the vertical direction.
The coating film is bounded by two parallel planes: the top free surface at $z = H$ and the bottom substrate at $z = 0$.
We describe the Brownian motion of the particles by the Langevin equations
and solve them numerically.
For the $i$-th particle,
the time evolution of the velocity $\boldsymbol{v}_i$ and the position $\boldsymbol{r}_i$ is expressed by~\cite{snap1, seg2, snap2, snap3}
\begin{align}
m \frac{\mathrm{d} \boldsymbol{v}_i}{\mathrm{d} t} &= -\zeta \boldsymbol{v}_i + \boldsymbol{F}_i^\mathrm{R} 
+ \sum_j (\boldsymbol{F}_{ij}^\mathrm{cnt} + \boldsymbol{F}_{ij}^\mathrm{DLVO}) 
+ \boldsymbol{F}_i^\mathrm{cpl}, \nonumber \\
\frac{\mathrm{d} \boldsymbol{r}_i}{\mathrm{d} t} &= \boldsymbol{v}_i
\label{e2-2-1}.
\end{align}

As the influence of the ambient liquid, 
the hydrodynamic drag $-\zeta \boldsymbol{v}_i$ and 
the random force $\boldsymbol{F}_i^\mathrm{R}$ due to thermal fluctuations are considered,
while hydrodynamic interactions among the particles are neglected. 
The hydrodynamic drag is expressed by Stokes' law as $\zeta = 3 \pi \eta d$.
The three components of the random force are given as the stochastic variables obeying 
independent Gaussian distributions satisfying
\begin{align}
\langle \boldsymbol{F}_i^\mathrm{R}(t) \rangle = \boldsymbol{0}, \ \ \ \ \
\langle \boldsymbol{F}_i^\mathrm{R}(t) \boldsymbol{F}_i^\mathrm{R}(0) \rangle 
= 2 k_\mathrm{B} T \zeta \delta(t) \boldsymbol{I}
\label{e2-2-2}.
\end{align}
The forces $-\zeta \boldsymbol{v}_i$ and $\boldsymbol{F}_i^\mathrm{R}$ cause the Brownian motion of the particle, 
whose diffusion coefficient is given by the Stokes--Einstein relation $D = k_\mathrm{B} T / \zeta$.

The contact force $\boldsymbol{F}_{ij}^\mathrm{cnt}$ describes collisions between the particles
and is given by the Voigt model with the Herzian contact theory:~\cite{Herz} 
\begin{align}
\boldsymbol{F}_{ij}^\mathrm{cnt} 
= \left[ \frac{1}{3} E^\ast d^{1/2} \delta_{ij}^{3/2} 
- \Gamma (\boldsymbol{v}_i - \boldsymbol{v}_j) \cdot \hat{\boldsymbol{n}}_{ij} \right] \hat{\boldsymbol{n}}_{ij}
\label{e2-2-3}.
\end{align}
The summation on the index $j$ includes the force from the substrate as well as the other particles.
The contact force acts in the direction normal to the tangent plane at the contact point.
This direction is expressed by the unit vector 
$\hat{\boldsymbol{n}}_{ij} = \boldsymbol{r}_{ij}/|\boldsymbol{r}_{ij}|$
with $\boldsymbol{r}_{ij} = \boldsymbol{r}_i - \boldsymbol{r}_j$.
The normal relative displacement of the contact point is given by
$\delta_{ij} = \max ( 0, d - |\boldsymbol{r}_{ij}| )$.
The longitudinal elastic modulus $E^\ast$ is given by the Young's modulus $E$ and the Poisson ratio $\nu$ as
$E^\ast = E/(1 - \nu^2)$.
We assume that the damping coefficient is related to the elastic modulus as
$\Gamma  = \lambda (m E^\ast d^{1/2} \delta_{ij}^{1/2} )^{1/2}$
to describe partially inelastic collisions.~\cite{cnt}
The contact force from the substrate is expressed by Eq.~(\ref{e2-2-3})
with substitution of $E^\ast \rightarrow 2^{1/2}E^\ast$ and $\delta_{ij} = \max ( 0, d/2 - z_i )$,
where $z_i = \hat{\boldsymbol{z}} \cdot \boldsymbol{r}_i$ is the $z$-coordinate of the $i$-th particle.
When an adhesive force $F_\mathrm{a}$ acts between the particles, 
there is an equilibrium displacement given by $\delta_0 = [ 3 F_\mathrm{a} /(E^\ast d^{1/2}) ]^{2/3}$.
Considering small oscillation around $\delta_0$, 
the elastic force is approximated by a linear restoring force with a spring constant of $E^\ast(d \delta_0)^{1/2}/2$.
Critical damping is realized when the factor of the damping coefficient $\Gamma$ satisfies $\lambda = 1$,
which we set in the present study.

The DLVO force $\boldsymbol{F}_{ij}^\mathrm{DLVO}$ describes the force acting between charged particles in liquid.
This force is given by the DLVO potential $V$ as~\cite{dlvo}
\begin{align}
\boldsymbol{F}_{ij}^\mathrm{DLVO} 
= \left. -\frac{\mathrm{d} V (h)}{\mathrm{d} h} \right|_{h = h_{ij}} \hat{\boldsymbol{n}}_{ij}
\label{e2-2-4},
\end{align}
\begin{align}
V(h)
= -\frac{Ad}{24 h}
+ \pi \varepsilon_\mathrm{r} \varepsilon_0 d \psi_\mathrm{eff}^2 e^{-\kappa h}
\label{e2-2-5},
\end{align}
where $h$ is the surface separation between the particles.
The DLVO potential is the combination of the van der Waals attraction (first term) and the electric double layer repulsion (second term).
The magnitude of the van der Waals attraction is provided by the Hamaker constant $A$.
The electric double layer repulsion is characterized by the effective surface potential $\psi_\mathrm{eff}$ and the Debye parameter $\kappa$,
and they are given by
\begin{align}
\psi_\mathrm{eff} = \frac{4 k_\mathrm{B} T}{Z \mathrm{e}} \tanh \left( \frac{Z  \mathrm{e} \psi_0}{4 k_\mathrm{B} T} \right), \ \ \ 
\kappa = \left( \frac{2 N_\mathrm{A} \mathrm{e}^2 I}{\varepsilon_\mathrm{r} \varepsilon_0 k_\mathrm{B} T} \right)^{1/2}
\label{e2-2-6},
\end{align}
where $\mathrm{e}$ is the elementary charge, $N_\mathrm{A}$ is the Avogadro constant, $\varepsilon_0$ is the electric constant,
$\varepsilon_\mathrm{r}$ is the relative permittivity, $Z$ is the ion valence, and $\psi_0$ is the surface electric potential.
The expression of $\psi_\mathrm{eff}$ in Eq.~(\ref{e2-2-6}) is that for valence symmetric electrolytes.
In this case, the ionic strength $I$ equals the electrolyte concentration multiplied by a factor of $Z^2$.
The inverse of the Debye parameter $\kappa^{-1}$ serves as a measure of the thickness of the electric double layer.
Note that the expression of Eq.~(\ref{e2-2-5}) is the approximate form for $\kappa^{-1} \ll h \ll d$, and we use it for simplicity.
We introduce a cut off distance $h_\mathrm{c}$ within which the DLVO force becomes constant,
and hence the adhesive force between the particles is given by $F_\mathrm{a} = (\mathrm{d} V/\mathrm{d} h)_{h = h_\mathrm{c}}$.
The cut off distance can be interpreted as the atomic length scale where the continuum picture, which is assumed in Eq.~(\ref{e2-2-5}), breaks down.~\cite{dlvo}
We calculate the DLVO force with setting $h_{ij} = \max (h_\mathrm{c}, |\boldsymbol{r}_{ij}|-d)$.

The capillary force $\boldsymbol{F}_i^\mathrm{cpl}$ acts to restore the particles
to the position satisfying an equilibrium contact angle with the free surface.
When the particles prefer to be perfectly wet, 
the capillary force is expressed as
\begin{align}
\boldsymbol{F}_i^\mathrm{cpl} = -4 \pi \gamma \delta_i^\mathrm{fs} \left( 1 - \frac{\delta_i^\mathrm{fs}}{d} \right)
\hat{\boldsymbol{z}}
\label{e2-2-7},
\end{align}
where $\gamma$ is the surface tension of the liquid.
The protruding length of the particle from the free surface is given by $\delta_i^\mathrm{fs} = \max [ 0, d/2 - (H - z_i) ]$.
To describe the evaporation of the liquid,
the free surface moves in the $-z$ direction with a rate of $U$.
The particles are swept by the free surface through the capillary force,
thereby forming concentrated particle layers. 
Although the ion concentration and the temperature can change during drying in real systems,
we assume that they are uniformly constant
and that the DLVO potential does not change.

According to Eq.~(\ref{e2-1-1}), the drying rate $U$ changes with reflecting the formation of concentrated particle layers.
We define the particle layer as the cluster of the particles moving with the free surface.
The clustering particles are detected as the particles contacting each other, i.e. $\delta_{ij} < 0$.
When one of the particles that compose the cluster satisfies $\delta_i^\mathrm{fs} < 0$, we regard the cluster as the particle layer. 
The permeation resistance of the particle layer is calculated by integrating the resistivity of the cross-section of the computational domain vertical to the $z$-axis:
\begin{align}
R = \int_0^H \rho(z) \mathrm{d}z, \ \ \ \ 
\rho(z) = \frac{80}{[D_\mathrm{H} (z)]^2} \frac{S}{S_\mathrm{f}(z)}
\label{e2-2-8},
\end{align}
where $S$ is the cross-sectional area of the computational domain, $S_\mathrm{f}$ is that of the liquid flow path, 
and $D_\mathrm{H}$ is the hydraulic diameter.
Since the cross-section of the flow path through the particle layer is quite complex shape,
the resistivity is evaluated by regarding the flow path as a cylindrical channel with a diameter of $D_\mathrm{H}$.~\cite{resist}
The evaluation of the resistivity in Eq.~(\ref{e2-2-8}) is an analogy to 
the resistivity of a cylindrical channel whose diameter is $D$, i.e. $32/D^2$, which is derived from the Hagen--Poiseuille equation.
The factor $S/S_\mathrm{f}$ appears to consider the variation of the flow rate inversely proportional to the cross-sectional area of the flow path.
The factor 80 reflects the channel tortuosity and is selected such that the Kozeny--Carman equation Eq.~(\ref{e2-1-12}) can be derived from Eq.~(\ref{e2-2-8})
as indicated below.
The hydraulic diameter $D_\mathrm{H}$ is given by the cross-sectional area of the flow path $S_\mathrm{f}$
and the wetted perimeter of the cross-section $L_\mathrm{f}$ as~\cite{resist}
\begin{align}
D_\mathrm{H}  = \frac{4 S_\mathrm{f}}{L_\mathrm{f}} 
\label{e2-2-9},
\end{align}
\begin{align}
S_\mathrm{f}(z)  = S - \frac{\pi}{4} d^2 \sum_{i \in P} \left[ 1 - \frac{4 (z - z_i)^2}{d^2} \right] \Theta \left( 1 - \frac{2|z - z_i|}{d} \right)
\label{e2-2-10},
\end{align} 
\begin{align}
L_\mathrm{f}(z)  = \pi d \sum_{i \in P} \Theta \left( 1 - \frac{2|z - z_i|}{d} \right)
\label{e2-2-11},
\end{align} 
where the Heaviside step function $\Theta$ appears to indicate the region where the particles occupy: $|z_i - z| \leq d/2$.
The index in the summation $i \in P$ indicates the particles that compose the particle layer.
Assuming the uniform distribution of the particles with a volume fraction of $\Phi_\mathrm{c}$,
we obtain $S_\mathrm{f} = (1 - \Phi_\mathrm{c})S$ and $L_\mathrm{f} = 6\Phi_\mathrm{c} S /d$, 
which lead to the Kozeny--Carman equation given by Eq.~(\ref{e2-1-12}).

\section{Results and discussion}

\begin{table}
\caption{
Time scales in the motion of particles.
}
\label{t1}
\centering
\scalebox{0.9}[0.9]{
\renewcommand\arraystretch{1.4}
\begin{tabular}{p{2.9cm}p{3.8cm}l}
\bhline{1.0pt} 
\multicolumn{2}{l}{\raisebox{0.em}{Time scale}} & 
\multicolumn{1}{l}{\raisebox{0.em}{Ratio to $\tau_\mathrm{hyd}$}} 
\\
\bhline{1.0pt} 
\multicolumn{3}{l}{Migration (Characteristic length: $d$)}\\
\hline
Evaporation & $\tau_\mathrm{E} = d/U_0$ & $ \ 25$\\
Diffusion & $\tau_\mathrm{D} = d^2/D$ & $ \ 1.0 \times 10^4 $\\
\hline
\multicolumn{3}{l}{Velocity change} \\
\hline
Hydrodynamic drag & $\tau_\mathrm{hyd} = m/\zeta$ & $ \ 1$\\ 
Contact force & $\tau_\mathrm{cnt} = (m/E^\ast)^{1/2} (d \delta_0)^{-1/4} $ & $ \ 1.0 \times 10^{-1}$\\
Capillary force & $\tau_\mathrm{cpl} = (m/\gamma)^{1/2}$ & $ \ 3.1 \times 10^{-1}$\\
\bhline{1.0pt}  
\end{tabular}
}
\end{table}

\begin{table}
\caption{
Dimensionless Debye parameter $\kappa d$ and potential barrier $\Delta E$ of DLVO potentials
for different ionic strengths. 
}
\label{t2}
\centering
\scalebox{0.9}[0.9]{
\renewcommand\arraystretch{1.4}
\begin{tabular}{p{1.8cm}rr}
\bhline{1.0pt} 
$I / (\mathrm{mol/L})$ & $\ \ \kappa d \ \ \ $ & $\Delta E/ (k_\mathrm{B} T) $\\
\bhline{1.0pt}
\ 0.01 & 6.5 \ \ \ & $ 15.8 \ \ \ $\\
\ 0.03 & 11.3 \ \ \ & $ 13.7 \ \ \ $\\
\ 0.1 & 20.5  \ \ \ & $ 10.7 \ \ \ $\\
\ 0.3 & 35.6 \ \ \ & $ 7.5 \ \ \ $\\
\ 1 & 64.9 \ \ \ & $ 3.5 \ \ \ $\\
\bhline{1.0pt}  
\end{tabular}
}
\end{table}

\begin{figure}
 \centering
 \includegraphics[width=7cm]{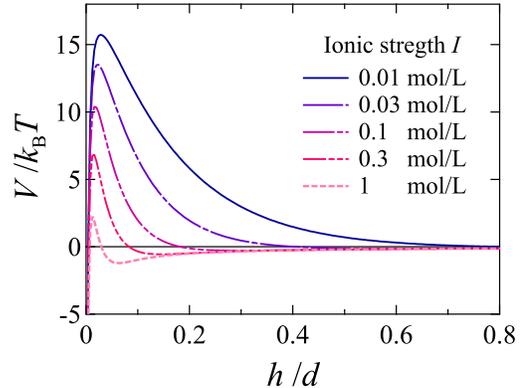}
 \caption{
DLVO potentials for different ionic strengths, as functions of surface separation between particles.
}
 \label{f4}
\end{figure}

\begin{figure}
 \centering
 \includegraphics[width=8cm]{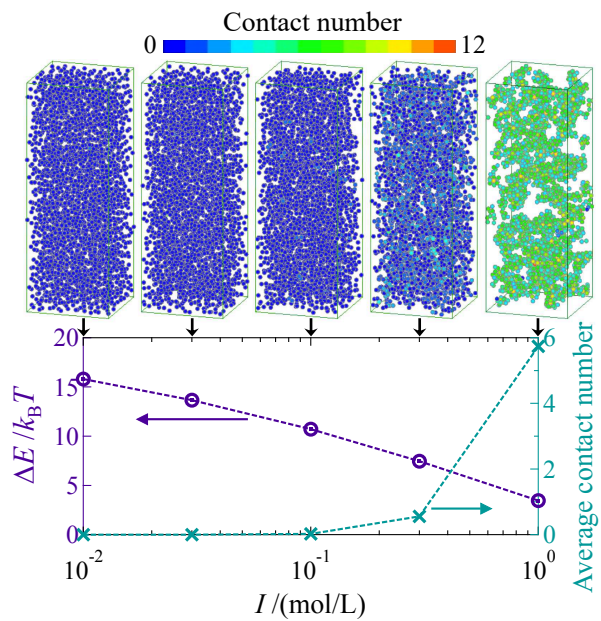}
 \caption{
Potential barrier (left axis) and the average contact number of particles for the initial configurations (right axis),
as functions of ionic strength. 
Corresponding snapshots of the initial configurations are shown in upper row, 
where the contact number of the particles is indicated by their color.
}
 \label{f5}
\end{figure}

Numerical simulations are performed with varying the DLVO potential.
We investigate the effects of the interaction between the particles on the drying characteristics.
To describe a part of the coating film far from edges,
periodic boundary conditions are applied in the directions parallel to the substrate with a side length of $18d$.
In the initial coating film of the thickness of $50d$, the volume fraction of the particles is set as $\Phi_0 = 0.1$.

The particle motion described by Eq.~(\ref{e2-2-1}) contains multiple time scales as listed on Table~\ref{t1}.
The time integration of Eq.~(\ref{e2-2-1}) is performed by use of the velocity Verlet algorithm.
The time increment is set as $0.2\tau_\mathrm{cnt}$ to be smaller than the shortest time scale in Table~\ref{t1}, i.e. $\tau_\mathrm{cnt}$.
The drying P\'{e}clet number is $\mathrm{Pe} = \tau_\mathrm{D}/\tau_\mathrm{E} = 400$,
which corresponds to $U_0 = 0.44 \ \mathrm{m/s}$ for an aqueous suspension with
$d = 20 \ \mathrm{nm}$, $\eta = 1 \times 10^{-3} \ \mathrm{Pa \ s}$, and $T = 300 \ \mathrm{K}$.
This drying rate is larger by a factor of about $10^6$ than actual situations.
The drying P\'{e}clet number represents also the ratio of the particle diameter to the diffusion length of the particles concentrated by the receding free surface,~\cite{clyr1,seg3}
 i.e. $\mathrm{Pe} = d/(D/U_0)$.
The current setting of $\mathrm{Pe} $ corresponds to reducing the length scale in the drying direction to perform simulations in smaller computational domains. 
We correspondingly set the characteristic thickness given in Sec.~\ref{s2-1} as $L_0 / d = 2$, 
which is about $10^{-6}$ times smaller than actual situations as demonstrated in Fig.~\ref{f3}.
The humidity is set as $\varphi = 0.5$. 

The dimensionless parameters of the DLVO potential are set as
$A/(24 k_\mathrm{B} T) = 0.1$ and $\pi \varepsilon_\mathrm{r} \varepsilon_0 d \Psi_\mathrm{eff}^2 /(k_\mathrm{B} T) = 23$.
This setting corresponds to considering the silica particles in water with 
$d = 20 \ \mathrm{nm}$, $T = 300 \ \mathrm{K}$, $\psi_0 = -50 \ \mathrm{mV}$, and $A = 1 \times 10^{-20} \ \mathrm{J}$.~\cite{hamak}
As shown in Table~\ref{t2} and Fig.~\ref{f4}, we change the other dimensionless parameter $\kappa d$ by changing the ionic strength according to Eq.~(\ref{e2-2-6}).
We define the potential barrier $\Delta E$ as the difference between the maximum and the local minimum (potential well) of the DLVO potential.
The potential barrier decreases with an increase in the ionic strength.
The cut off length $h_\mathrm{c}$ is set to satisfy $\delta_0 /d = 0.02$.
In the present simulations, we consider the DLVO force only between the particles with the separation of $h/d \leq 1$,
where the DLVO potential sufficiently decays since all the settings in Table~\ref{t2} satisfies $(\kappa d)^{-1} < 1$. 
The initial configurations exhibited in Fig.~\ref{f5} are obtained as follows:
the particles are firstly distributed randomly so that the surface separations between the particles become more than that of the potential maximum,
and then the Brownian motion of the particles are solved during a time of $5 \tau_\mathrm{D}$.
The particles aggregate at $I \geq 0.3 \ \mathrm{mol/L}$ where $\Delta E/(k_\mathrm{B} T) \lesssim 10$ is satisfied.
The average contact number of the particles rises sharply accompanied by the aggregation.

\begin{figure}
 \centering
 \includegraphics[width=8cm]{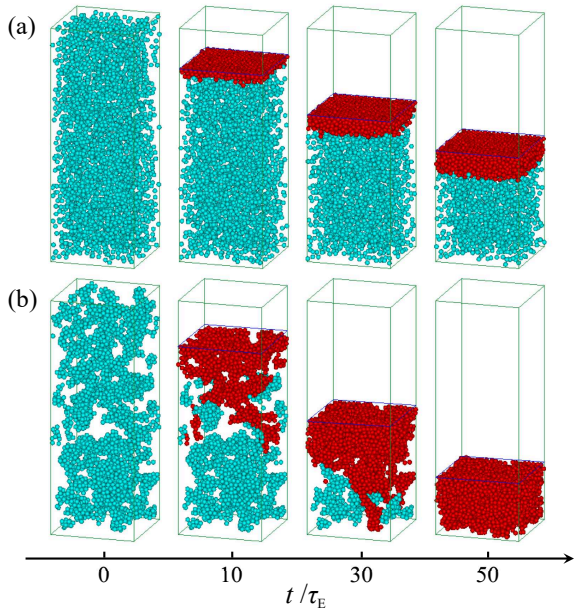}
 \caption{
Snapshots of the particle configurations during drying.
The ionic strengths are (a) $0.1 \ \mathrm{mol/L}$ and (b) $1 \ \mathrm{mol/L}$.
The particles composing the concentrated layers are indicated by the red color.
}
 \label{f6}
\end{figure}

\begin{figure}
 \centering
 \includegraphics[width=7cm]{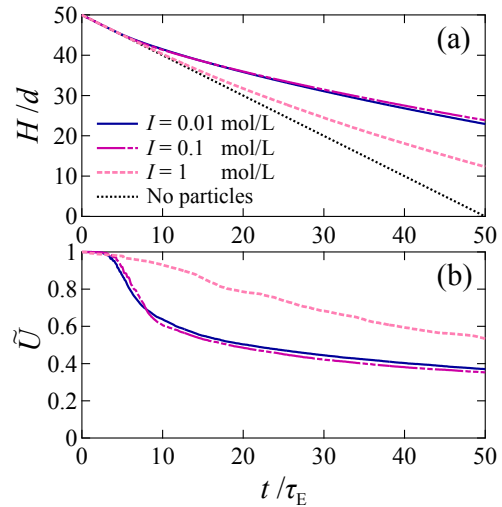}
 \caption{
Time variation of (a) coating film thickness and (b) drying rate for different ionic strengths.
The dotted line represents the time change of the film thickness containing no particles.
}
 \label{f7}
\end{figure}

As shown in Fig.~\ref{f6},
the receding free surface sweeps the particles to form the concentrated particle layers.
Figure~\ref{f7} thus indicates that the drying rate decreases with the growth of the particle layers.
If there are no particles, the thickness of the liquid film decreases linearly with a constant drying rate.
The drying rate decreases most slowly at the highest ionic strength of $I = 1 \ \mathrm{mol/L}$ where the particles aggregate.
Comparing the cases of $I \leq 0.1 \ \mathrm{mol/L}$ where the particles are stably dispersed,
the drying rate is initially larger at the higher ionic strength of $I = 0.1\ \mathrm{mol/L}$,
while it finally becomes larger at the lower ionic strength of $I = 0.01\ \mathrm{mol/L}$.
This difference reflects the permeation resistance as discussed below.

\begin{figure}
 \centering
 \includegraphics[width=9cm]{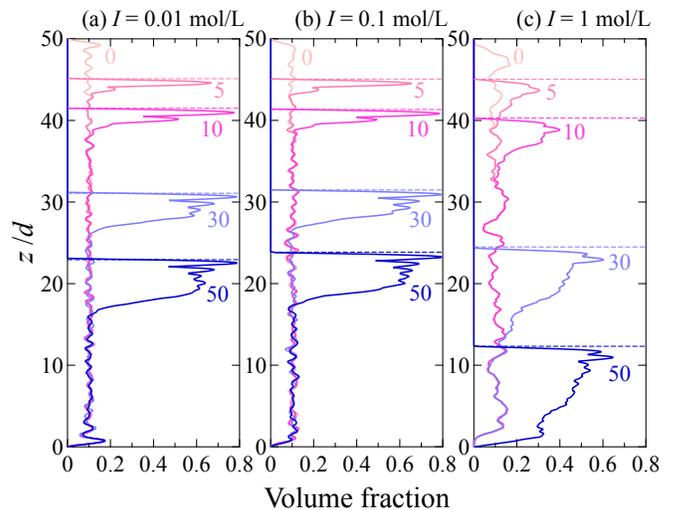}
 \caption{
Distributions of particle volume fraction at the times of $t/\tau_\mathrm{E} = 0, 5, 10, 30, 50$
for different ionic strengths.
The broken lines represent the position of the free surface. 
}
 \label{f8}
\end{figure}

Figure~\ref{f8} demonstrates the growth of the particle layers as the increase in the particle volume fraction beneath the free surface,
where the clear oscillatory profiles can be found except at $I = 1 \ \mathrm{mol/L}$.
This oscillatory profile indicates the stratification of the particles.
The DLVO potential barrier hinders the particles to contact each other,
thereby allowing the concentrated particles to rearrange to stratify.
When the potential barrier is too low so that the particles aggregate at $I = 1 \ \mathrm{mol/L}$,
the particles cannot easily rearrange and form loose particle layers,
thus the oscillatory profiles do not appear.

\begin{figure}
 \centering
 \includegraphics[width=7cm]{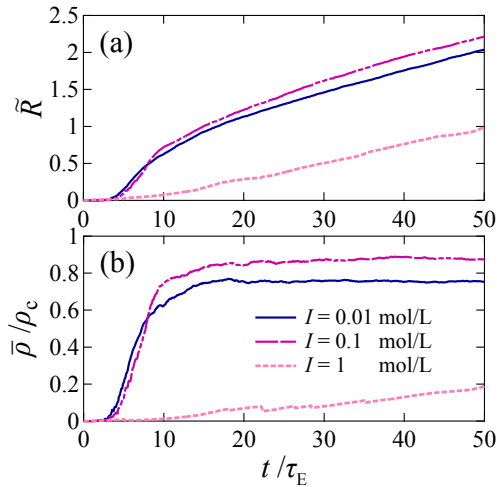}
 \caption{
Time variation of (a) the permeation resistance and (b) the resistivity of concentrated particle layers for different ionic strengths.
The resistivity is scaled by that evaluated by the Kozeny--Carman equation Eq.~(\ref{e2-1-12}) with $\Phi_\mathrm{c} = 0.64$.
}
 \label{f9}
\end{figure}

The permeation resistance increases with the growth of the concentrated particle layers as shown in Fig.~\ref{f9}(a).
The increase of the permeation resistance results in the decrease of the drying rate through Eq.~(\ref{e2-1-4}).
The larger permeation resistance is resulted from the larger particle volume fraction of the particle layers.
In the cases of $I \leq 0.1 \ \mathrm{mol/L}$, 
the permeation resistance increases sharply at $t/\tau_\mathrm{E} \approx 5$,
which seems to correspond to the time when the first layer of the particle layer is formed.
In fact, this time is roughly evaluated as $t/\tau_\mathrm{E} = (\Phi_\mathrm{c} - \Phi_0)/\Phi_0$,
which equals $5.4$ with assuming $\Phi_\mathrm{c} = 0.64$.
This evaluation is derived from the particle mass balance of $(\Phi_\mathrm{c} - \Phi_0) d = \Phi_0 U_0 t$.
Comparing the cases of $I \leq 0.1 \ \mathrm{mol/L}$,
the permeation resistance is initially larger for the smaller ionic strength of $I = 0.01 \ \mathrm{mol/L}$.
In Fig.~\ref{f8}, the initial distributions at $t/\tau_\mathrm{E} = 0$ show that
the stronger repulsive interaction between the particles results in the larger particle volume fraction beneath the free surface,
and thus the permeation resistance becomes larger.
Such increased accumulation of particles near a wall by repulsive interactions between particles was also shown in the previous study.~\cite{wall}
However, the permeation resistance finally becomes larger at the larger ionic strength of $I = 0.1 \ \mathrm{mol/L}$,
corresponding to the larger particle volume fraction of the particle layer as demonstrated by the profiles at $t/\tau_\mathrm{E} \geq 10$ in Fig.~\ref{f8}.
This larger particle volume fraction can be attributed to the smaller potential barrier and potential range, which would allow the particles to be densely packed.

In Fig.~\ref{f9}(b), the average resistivity is evaluated as $\bar{\rho} = R / [2 (H - \langle z \rangle_P )]$, 
where $\langle z \rangle_P$ is the mean $z$-coordinate of the particles that compose the particle layers,
and $2 (H - \langle z \rangle_P )$ gives an estimation of the thickness of the particle layers.
The resistivity also increases with time but converges in the cases of the stably dispersed particles at $I \leq 0.1 \ \mathrm{mol/L}$.
Although the particle volume fraction of the particle layer is not uniform especially near the free surface as shown in Fig.~\ref{f8},
the convergence of the resistivity suggests that the particle layer gradually become uniform with distance from the free surface.
However, such convergence is not observed at $I = 1 \ \mathrm{mol/L}$,
where the particle volume fraction of the particle layer continuously increases with time.
When we can observe a convergence of the resistivity in the Langevin dynamics simulations,
combining this evaluated resistivity and the continuum model given in \ref{sA-2}
would enable us to evaluate the drying characteristics of coating films whose thickness is larger than the simulation scale.

\begin{figure}
 \centering
 \includegraphics[width=8cm]{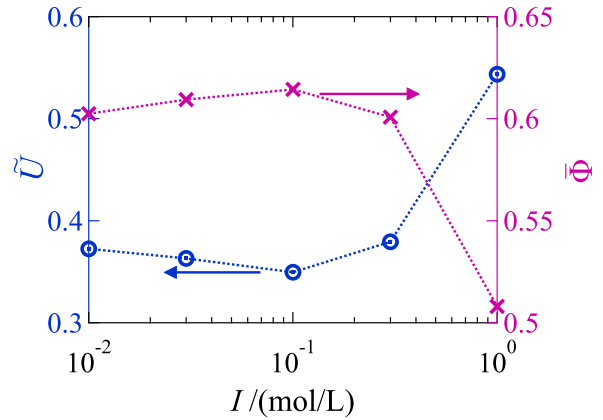}
 \caption{
Drying rate and the particle volume fraction of concentrated particle layers at $t / \tau_\mathrm{E} = 50$, as functions of ionic strength.
}
 \label{f10}
\end{figure}

Figure~\ref{f10} summarizes the effect of the interaction between the particles on the drying characteristics.
The particle volume fraction of the particle layer $\bar{\Phi}$ is evaluated 
as the mean particle volume fraction in the region within the depth of $3d$ from the free surface, i.e. $[H-3d, H]$.
As long as the particles stably dispersed at $I \leq 0.1 \ \mathrm{mol/L}$,
decreasing the potential barrier by increasing the ionic strength results in
a little decrease in the drying rate with a little increase in the particle volume fraction of the particle layer.
On the other hand, when the particles aggregate at the higher ionic strength of $I > 0.1 \ \mathrm{mol/L}$, 
we find obvious change that the particle layer becomes loose and the drying rate increases.
In the previous experimental study of drying of colloidal suspension droplets,
it was also indicated that the morphology of the dried grains can be changed 
by forming aggregates, due to the decrease in the permeation resistance of the particle layers.~\cite{sdry3}

\section{Conclusions}

We constructed a model that can calculate the drying characteristics by use of Langevin dynamics simulations.
This model enables us to evaluate the drying characteristics of colloidal suspensions from the character of colloidal particles such as interactions.
The present calculation of this model shows that the formation of aggregates by attractive interactions suppresses the decrease in the drying rate.
This suppression is due to the loose structure of the concentrated particle layers.
The aggregation is thus preferable to reduce the drying time,
while the final dried structure would be loose and disordered.~\cite{snap3}
To improve the density and order of structures with reducing the drying time by aggregation,
further investigations considering the effects of morphology and strength of the aggregates would be required.
Note that the present finding will be true for filtration.
The aggregated particles are expected to form loose cake, thereby suppressing the decrease in the filtering flux.
Both filtration and drying are operations of solid--liquid separation where liquid is removed through an interface.

\appendix
\renewcommand{\thesection}{Appendix}
\renewcommand{\thesubsection}{\Alph{subsection}}
\renewcommand{\theequation}{A\arabic{equation}}
\renewcommand{\thefigure}{A\arabic{figure}}
\setcounter{equation}{0}
\setcounter{figure}{0}
\section*{\label{sA}Appendix}

\subsection{\label{sA-1}Formulation of drying rates in terms of chemical potential}

\begin{figure}
 \centering
 \includegraphics[width=8cm]{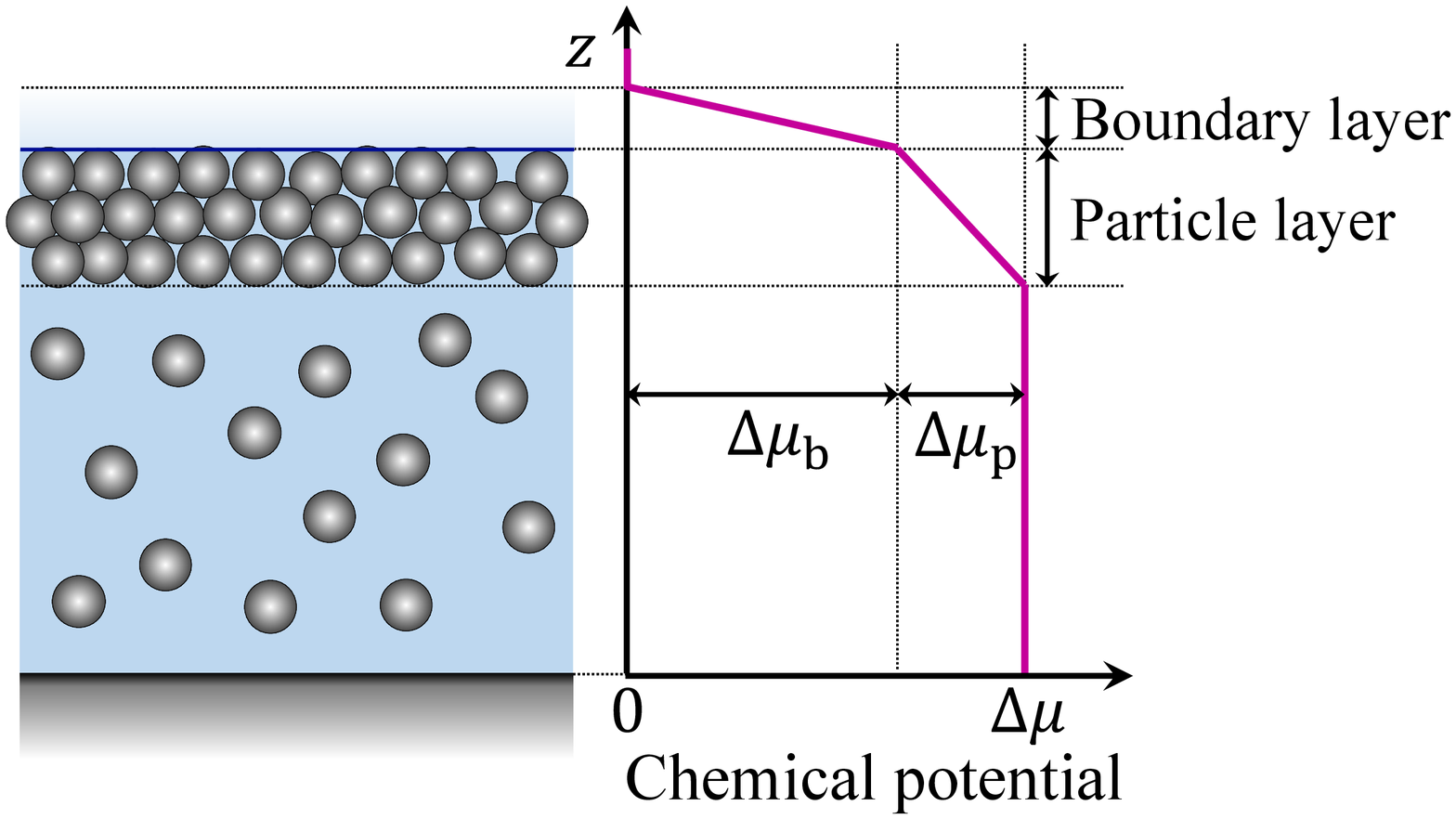}
 \caption{
Sketch of the chemical potential profile of the liquid.
The liquid-to-vapor phase transition occurs at the free surface.
The chemical potential drops in two steps:
permeation through the concentrated particle layer ($\Delta \mu_\mathrm{p}$)
and diffusion through the boundary layer ($\Delta \mu_\mathrm{b}$).
}
 \label{fA1}
\end{figure}

The evaporation can also be considered to be driven by the difference in chemical potential, 
instead of the difference in vapor pressure as indicated in Eq.~(\ref{e2-1-1}).
As shown in Fig.~\ref{fA1},
when a concentrated particle layer is formed beneath the free surface, 
the chemical potential of the liquid drops in two steps.
The total chemical potential drop is $- k_\mathrm{B} T \ln \varphi$
that equals the chemical potential difference between the vapor of the pressure $p_0$ and $p_\infty$:
\begin{align}
\Delta \tilde{\mu}_\mathrm{b} + \Delta \tilde{\mu}_\mathrm{p}
= \Delta \tilde{\mu} =  -\ln \varphi
\label{eA1-1},
\end{align}
where the chemical potential is scaled by $k_\mathrm{B} T$ and dimensionless quantities are denoted by the tilde.
The chemical potential drops through the boundary layer and the particle layer are given as
\begin{align}
\Delta \tilde{\mu}_\mathrm{b} =  \ln \frac{\tilde{p}}{\varphi}
\label{eA1-2},
\end{align}
\begin{align}
\Delta \tilde{\mu}_\mathrm{p} = \frac{V_\mathrm{m} \Delta P}{k_\mathrm{B} T}
\label{eA1-3},
\end{align}
respectively, where we denote $\tilde{p} = p/p_0$.
The drying rate is expressed to be proportional to the chemical potential drops in each step as
\begin{align}
\tilde{U} = \Lambda_\mathrm{b}(\tilde{p}) \Delta \tilde{\mu}_\mathrm{b}
= \Lambda_\mathrm{p} \Delta \tilde{\mu}_\mathrm{p}
\label{eA1-4}.
\end{align}
Comparing Eq.~(\ref{eA1-4}) with Eqs.~(\ref{e2-1-1}) and (\ref{e2-1-3}), the transport coefficients are expressed as
\begin{align}
\Lambda_\mathrm{b}(\tilde{p}) = \frac{\tilde{p} - \varphi}{(1 - \varphi) \ln (\tilde{p} /\varphi)}
\label{eA1-5},
\end{align}
\begin{align}
\Lambda_\mathrm{p} = \frac{1}{(1 - \varphi) \tilde{R}}
\label{eA1-6}.
\end{align}
Note that the former depends on the vapor pressure at the free surface $p$.
Combining Eqs.~(\ref{eA1-1}) and (\ref{eA1-4}),
we finally obtain an expression of the relationship between the drying rate and the total chemical potential drop as
\begin{align}
\tilde{U} = \left[ \frac{1}{\Lambda_\mathrm{b}(\tilde{p})} + \frac{1}{\Lambda_\mathrm{p}} \right]^{-1} \Delta \tilde{\mu}
\label{eA1-7}.
\end{align}
In this expression, the coefficient of $\Delta \tilde{\mu}$ represents the overall transport coefficient.
In the limit of $\varphi \rightarrow 1$, $\Lambda_\mathrm{b} \sim (1 - \varphi)^{-1}$ and $\Delta \tilde{\mu} \sim (1 - \varphi)$
result in Eq.~(\ref{e2-1-11}) that indicates a flow rate through serial resistance.

\subsection{\label{sA-2}Drying characteristics associated with the growth of uniform particle layers}

\begin{figure}
 \centering
 \includegraphics[height=6cm]{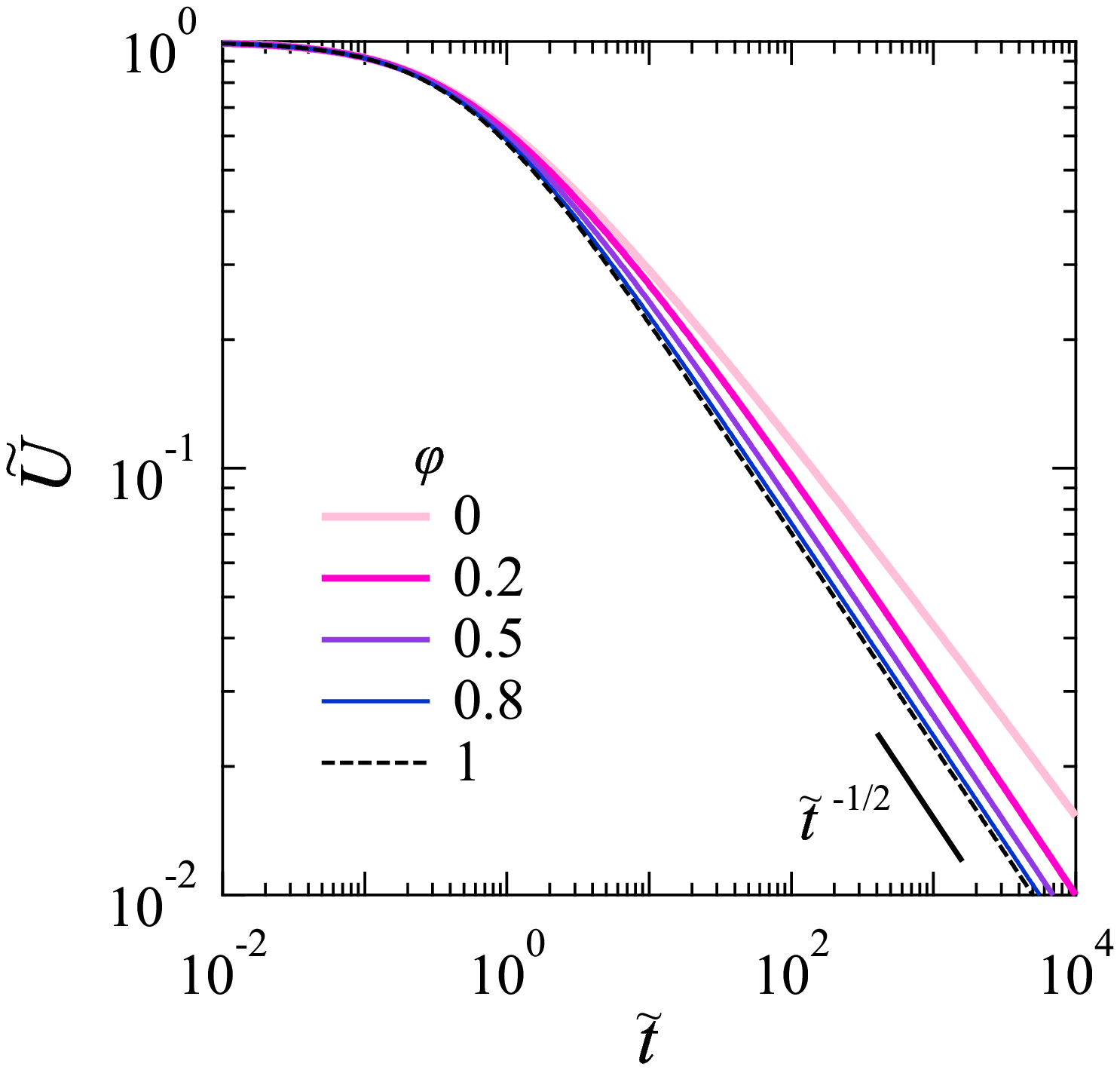}
 \caption{
Time variation of drying rate at different humidities.
The profiles are given by Eq.~(\ref{eA2-5}) for $0 < \varphi < 1$;
Eq.~(\ref{eA2-9}) for $\varphi = 0$; 
and Eq.~(\ref{eA2-11}) for $\varphi \rightarrow 1$.
}
 \label{fA2}
\end{figure}

\begin{figure}
 \centering
 \includegraphics[height=6cm]{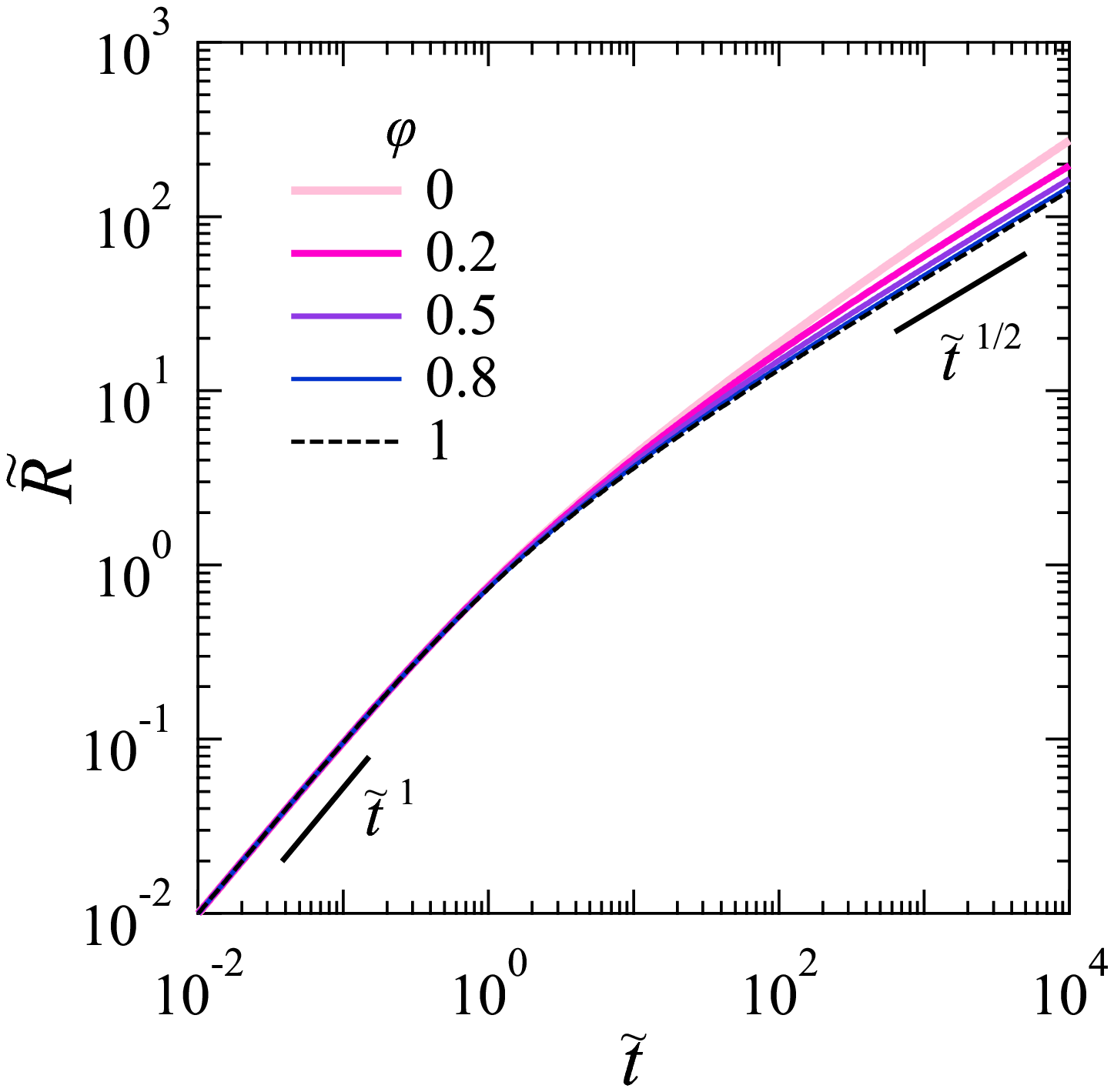}
 \caption{
Time variation of permeation resistance at different humidities.
This describes the growth of the concentrated particle layer.
The profiles are given by combining Eqs.~(\ref{e2-1-4}) and (\ref{eA2-5}) for $0 < \varphi < 1$;
combining Eqs.~(\ref{e2-1-4}) and (\ref{eA2-9}) for $\varphi = 0$; 
and Eq.~(\ref{eA2-12}) for $\varphi \rightarrow 1$
}
 \label{fA3}
\end{figure}

Assuming that the density of the particle layer is spatially uniform and does not change in time,  
the growth of the particle layer can be described by a one-dimensional differential equation as
\begin{align}
\frac{\mathrm{d} L}{\mathrm{d} t} = \alpha U, \ \ \ \ \
\alpha = \frac{\Phi_0}{\Phi_\mathrm{c} - \Phi_0}
\label{eA2-1}.
\end{align}
This equation is derived from the particle mass balance: $(\Phi_\mathrm{c} - \Phi_0) \mathrm{d}L = \Phi_0 U \mathrm{d}t$.
With the relation $R = \rho_\mathrm{c} L$,
a dimensionless form of Eq.~(\ref{eA2-1}) is obtained as 
\begin{align}
\frac{\mathrm{d} \tilde{R}}{\mathrm{d} \tilde{t}} = \tilde{U}
\label{eA2-2},
\end{align}
where the time is scaled as
\begin{align}
\tilde{t} = \frac{t}{t_0}, \ \ \ \ \
t_0 = \frac{L_0}{\alpha U_0}
\label{eA2-3}.
\end{align}
An explicit expression of $\tilde{R}$ as a function of $\tilde{U}$ is obtained from Eq.~(\ref{e2-1-4}).
Substituting this explicit expression to Eq.~(\ref{eA2-2}) yields the following differential equation:
\begin{align}
\frac{\mathrm{d} \tilde{U}}{\mathrm{d} \tilde{t}} 
= \tilde{U}^3 \left\{ \frac{\ln[(1 - \varphi)\tilde{U} + \varphi]}{1 - \varphi} - \frac{\tilde{U}}{(1 - \varphi)\tilde{U} + \varphi} \right\}^{-1}
\label{eA2-4}.
\end{align}
The solution of Eq.~(\ref{eA2-4}) is obtained as
\begin{align}
\tilde{t}
= &-\frac{\ln[(1 - \varphi)\tilde{U} + \varphi]}{2(1 - \varphi) \tilde{U}^2} 
+ \frac{1}{2 \varphi \tilde{U}} \nonumber \\
&- \frac{1 - \varphi}{2 \varphi^2} \ln \left[ 1 + \frac{\varphi}{(1 - \varphi) \tilde{U}} \right]
-\frac{1}{2 \varphi} \left[ 1 + \frac{1 - \varphi}{\varphi} \ln(1 - \varphi) \right]
\label{eA2-5}.
\end{align}
As shown in Figs.~\ref{fA2} and \ref{fA3}, 
the drying rate decreases with time,
and thus the increasing rate of the permeation resistance is reduced.
Unlike the simulation results of $I  \leq 0.1 \ \mathrm{mol/L}$ in Fig.~\ref{f9}(a), sharp increases at an early stage do not appear.
This is because the model in this section assumes 
the growth of spatially uniform particle layers even when their thickness is less than the particle diameter.
The influence of the humidity does not appear in the initial behavior of $\tilde{R} \sim \tilde{t}$ as $\tilde{t} \rightarrow 0$,
but it appears with time.
The long-time asymptotic behavior for $\varphi > 0$ is
\begin{align}
\tilde{U} \sim \left[ \frac{-\ln \varphi}{2(1 - \varphi)} \right]^{1/2} \tilde{t}^{-1/2}
\ \ \ \ (\tilde{t} \rightarrow \infty)
\label{eA2-6},
\end{align}
\begin{align}
\tilde{R} \sim \left( \frac{-2 \ln \varphi}{1 - \varphi} \right)^{1/2} \tilde{t}^{1/2}
\ \ \ \ (\tilde{t} \rightarrow \infty)
\label{eA2-7},
\end{align}
while a different asymptotic behavior is derived for $\varphi = 0$ as follows:
\begin{align}
\frac{\mathrm{d} \tilde{U}}{\mathrm{d} \tilde{t}} 
= \frac{\tilde{U}^3}{\ln \tilde{U} - 1}
\label{eA2-8},
\end{align}
\begin{align}
\tilde{t} = \frac{1}{4} \left( \frac{1 - 2 \ln \tilde{U}}{\tilde{U}^2} - 1 \right)
\sim -\frac{1}{2} \frac{\ln \tilde{U}}{\tilde{U}^2}
\ \ \ \ (\tilde{t} \rightarrow \infty)
\label{eA2-9}.
\end{align}
This expression indicates that the drying rate decreases more slowly than in the cases of $\varphi > 0$.

In the limit of $\varphi \rightarrow 1$,
Eq.~(\ref{eA2-4}) becomes a simple form as
\begin{align}
\frac{\mathrm{d} \tilde{U}}{\mathrm{d} \tilde{t}} 
= -\tilde{U}^3
\label{eA2-10}.
\end{align}
The solution of Eq.~(\ref{eA2-10}) is obtained as
\begin{align}
\tilde{U} = (1 + 2\tilde{t})^{-1/2}
\label{eA2-11},
\end{align}
\begin{align}
\tilde{R} = (1 + 2\tilde{t})^{1/2} - 1
\label{eA2-12}.
\end{align}
This result corresponds to the Ruth's filtration equation
that describes the temporal variation of the flow rate in dead-end filtration.~\cite{filt}
The Ruth's filtration equation is derived from Darcy's law with considering the series permeation resistance of the filter media and the cake deposited onto the filter.
In the limit of $\varphi \rightarrow 1$, the present model also becomes an expression indicating series resistance as Eq.~(\ref{e2-1-11}).

In the previous experiments of unidirectional drying in confined systems,~\cite{udry1, udry2, udry3, udry4, udry5}
the change in the time dependence of the particle layer thickness from $t$ to $t^{1/2}$ was observed.
To explain this behavior, equations that have the same form as Eq.~(\ref{eA2-12}) were suggested,
but the derivation of them was empirical or approximative.~\cite{udry1, udry5}
With considering the influence of humidity, which was ignored in the previous studies,~\cite{udry1, udry5}
the present model yields the analytical expressions of the drying characteristics
and shows that the previously suggested expressions hold only for the limit of $\varphi \rightarrow 1$.
Figures~\ref{fA2} and \ref{fA3} shows that the humidity affects the drying characteristics especially for long-time behavior.
The present model would be helpful to explain the drying characteristics observed in the experiments.

\section*{Acknowledgements}
This work was carried out under the project of Products Innovation Association, ``Structure of NAno Particles (SNAP) study group,''
and was supported by the Japan Society for the Promotion of Science (JSPS) KAKENHI Grant No. 19K15335 and 21K04747.

\end{document}